\documentclass{aa}  
\usepackage{natbib}
\bibpunct{(}{)}{;}{a}{}{,}
\usepackage{graphicx}
\usepackage{txfonts}
\begin{document}
   \title{The distance and neutral environment of the massive stellar
   cluster Westerlund~1}

   \author{R. Kothes
          \inst{1,2}
          \and
          S. M. Dougherty\inst{1}}
          
\titlerunning{Distance and Environment of Westerlund 1}
\authorrunning{Kothes \& Dougherty}

   \offprints{R. Kothes}

   \institute{National Research Council of Canada,
              Herzberg Institute of Astrophysics,
              Dominion Radio Astrophysical Observatory,
              P.O. Box 248, Penticton, British Columbia,
              V2A 6J9, Canada
         \and Department of Physics and Astronomy, University of
              Calgary, 2500 University Drive N.W., Calgary, AB,
              Canada
              }

   \date{Received 19 February 2007 / Accepted 13 April 2007}

 
  \abstract 
   {In spite of a large number of recent publications about the
   massive stellar cluster Westerlund~1, its distance from the Sun
   remains uncertain with values as low as 1.1 kpc, but largely between
   4 and 5 kpc.}
   {The goal of this study is to determine a distance to Westerlund~1 
   independent of the characteristics of the stellar population and to 
   study its neutral environment, using observations of atomic hydrogen.}
   {The \ion{H}{i} observations are taken from the Southern
   Galactic Plane Survey to study \ion{H}{i} absorption in the
   direction of the \ion{H}{ii} region created by the members of
   Westerlund~1 and to investigate its environment as observed in the
   \ion{H}{i} line emission. A Galactic rotation curve was derived using 
   the recently revised values for the Galactic centre distance of 
   $R_\odot = 7.6$~kpc, and the velocity of the 
   Sun around the Galactic centre of $\Theta_\odot = 214$~km\,s$^{-1}$. 
   This rotation curve successfully predicts the location of the Tangent point gas
   and the velocity of the Sagittarius Arm outside the solar circle on
   the far side of the Galaxy to within 4 km~s$^{-1}$.
   Compared to the typically used values of $R_\odot = 8.5$~kpc and 
   $\Theta_\odot = 220$~km\,s$^{-1}$ this reduces kinematically determined 
   distances by more than 10~\%.}
   {The newly determined rotation
   model leads us to derive a distance of $3.9\pm 0.7$~kpc to
   Westerlund~1, consistent with a location in the Scutum--Crux
   Arm. Included in this estimate is a very careful investigation
   of possible sources of error for the Galactic rotation curve. 
   We also report on small expanding \ion{H}{i} features around
   the cluster with a maximum dynamic age of 600,000~years and a
   larger bubble which has a minimum dynamic age of 2.5 million
   years. Additionally we re-calculated the kinematic distances to
   nearby \ion{H}{ii} regions and supernova remnants based on
   our new Galaxic rotation curve.}
   {We propose that in the early stages of the development of Wd~1 a
   large interstellar bubble of diameter about 50~pc was created by
   the cluster members. This bubble has a dynamic
   age similar to the age of the cluster. Small expanding bubbles,
   with dynamical ages $\sim0.6$~Myr are found around Wd~1, which we
   suggest consist of recombined material lost by cluster members
   through their winds.}


   \maketitle
%

\section{Introduction}

Westerlund~1 (Wd~1) is a highly reddened compact cluster with a large
population of post-main sequence massive stars, including OB
supergiants and hypergiants, red and yellow supergiants, and
Wolf-Rayet (WR) stars \citep{clar02,clar05}.  The mass of Wd~1 is
likely to be in excess of 10$^5$M$_{\odot}$ \citep{clar05}, exceeding
that of any of the other known massive Galactic clusters, including
NGC 6303 \citep{crow98}, the Arches \citep{fige02} and Quintuplet
\citep{fige99}, and is more comparable to the mass of Super Star
Clusters (SSC), previously identified only in other galaxies.  If Wd~1
is indeed a Super Star Cluster within our own Galaxy \citep{clar05},
this is a unique opportunity to study the properties of a nearby SSC,
where it is possible to resolve the individual massive stars, and
determine basic properties more readily than in the typically more
distant examples.

In spite of a large number of recent observations of Wd~1, its
distance from the Sun remains somewhat uncertain. \citet{piat98}
derived a distance of $1.1\pm0.4$~kpc based on photometry of the OB
supergiants identified at the time.  Using the yellow hypergiants
where a spectroscopic luminosity discriminant is available that is
lacking for OB supergiants, \citet{clar05} determined an upper limit
for the distance of 5.5~kpc. \citet{clar05} also argue for a lower
limit on the distance of 2~kpc based on an analysis of radio continuum
data \citep{clar98}. They also argue that the
comparatively low distance of \citet{piat98} is the result of an error
in both the absolute magnitude calibration and the reddening law that
was used.  Additionally, \citet{clar05} did not identify {\em any}
dwarf or giant stars, with all the OB stars being supergiants, unlike
the work of Piatti et al. where detection of the main sequence was
claimed.  Further support for a distance in excess of 2~kpc comes from
analysis of near-IR photometry of the WR stars in Wd~1 \citep{crow06},
where the average distance modulus derived from 23 of the 24 known WR
stars leads to a distance of $4.7\pm1.1$~kpc. The relatively large
uncertainty in this estimate is a result of the uncertainty in the
absolute magnitude calibration of WR stars. A more precise estimate of
distance comes from an initial analysis of deep IR imaging with the
VLT that reveals the main sequence and pre-main sequence populations
in Wd~1 from which a photometric distance of $4.0\pm0.3$~kpc is
deduced \citep{bran05}.

The most recent analyses appear to be converging on a distance in the
range 4-5~kpc. In this paper, we examine \ion{H}{i} data in the
direction of Wd~1 from the Southern Galactic Plane Survey (SGPS)
\citep{mccl05} to estimate a distance to Wd~1 based on \ion{H}{i}
absorption and related \ion{H}{i} features. We also examine the
large-scale distribution of \ion{H}{i} in the environment of Wd~1 to
search for evidence of the impact of the putative SSC on the
surrounding interstellar medium.

\section{The SGPS \ion{H}{i} data set}

The Southern Galactic Plane Survey \citep{mccl05} is part of the
International Galactic Plane Survey (IGPS), a project devoted to
mapping the neutral hydrogen in the plane of the Galaxy,
the Milky Way, at arcminute scale. The SGPS maps the region of the Milky Way
visible in the southern hemisphere. Other parts of 
the IGPS are the Canadian Galactic Plane Survey \citep[CGPS,][]{tayl03},
which covers the Galactic plane seen in the northern hemisphere,
and the VLA Galactic Plane Survey \citep[VGPS,][]{stil06}, which 
connects the two other surveys through the first quadrant of the
Milky Way.

The SGPS data were obtained with the Australia Telescope Compact Array
(ATCA) and the Parkes 64-m radio telescope. The survey covers an area
of $325\degr^2$ from $253\degr$ to $358\degr$ and $5\degr$ to
$20\degr$ in Galactic longitude and from $-1\fdg5$ to $+1\fdg5$ in
Galactic latitude. In the direction of Wd~1 the total velocity
coverage is $600$~km~s$^{-1}$ from $-300$ to $+300$~km~s$^{-1}$. The
angular resolution is $\sim2\arcmin$ and the sensitivity $\sim1.6$~K.

\section{Results}

\subsection{Galactic kinematics in the direction of Westerlund~1}

\begin{figure}
   \centerline{\includegraphics[bb = 50 45 530 500,width=8.5cm,clip]{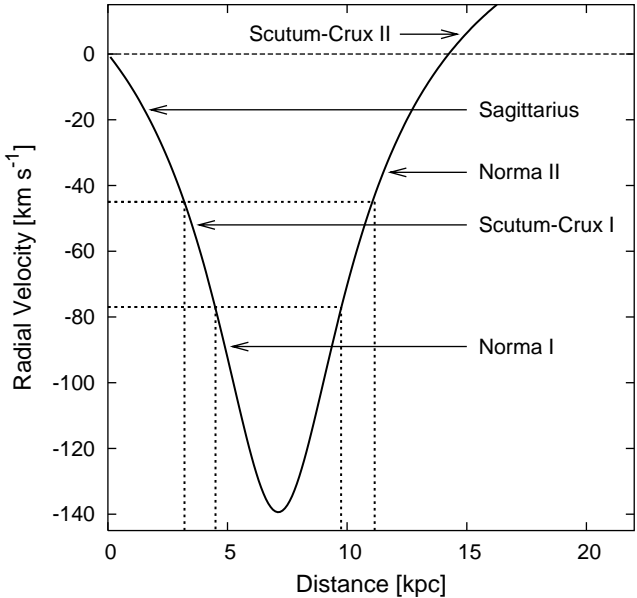}}
   \caption{The Galactic rotation curve toward Galactic longitude of $339\fdg6$
   using a Galactocentric radius of R$_\odot = 7.6$~kpc and a circular velocity
   of $\Theta_\odot = 214$~km~s$^{-1}$ for the Sun. We assumed a flat rotation curve and
   purely circular rotation. The approximate locations of the spiral arms in
   radial velocity are indicated. Here the indices I and II indicate the first
   and second time the line of sight passes through that spiral arm
   (see also Fig.~\ref{mw}).}
   \label{rot}
\end{figure}

\begin{figure}
   \centerline{\includegraphics[bb = 45 36 493 509,width=8.5cm,clip]{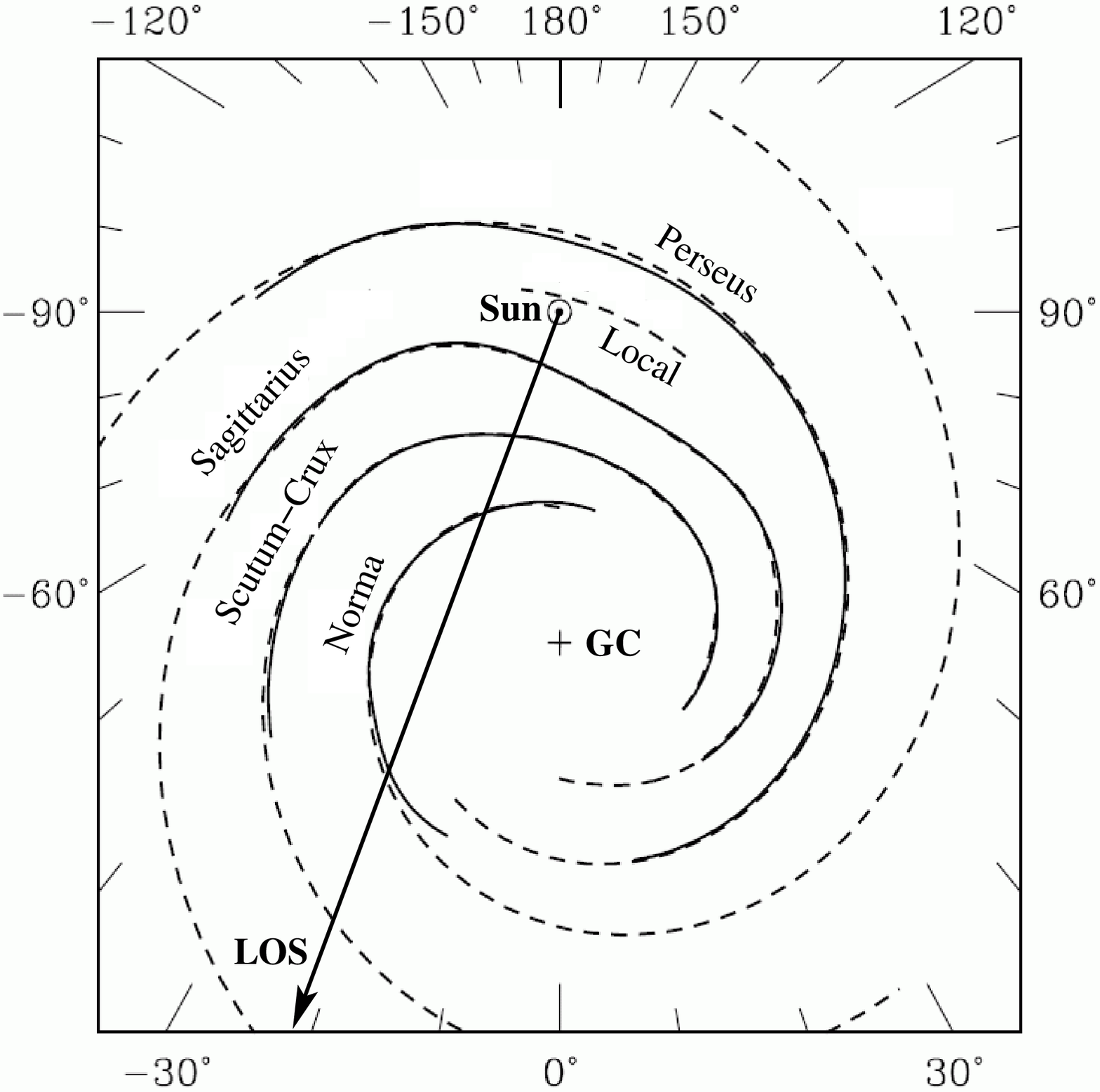}}
   \caption{The spiral arm pattern of the Milky Way Galaxy \citep{tayl93,cord02}.
   The names of the spiral arms and the line of sight (LOS) toward
   Wd~1 are indicated.}
   \label{mw}
\end{figure}

We carefully investigated the dynamics along the line of sight
towards Wd~1. As a starting point we used a
flat rotation model to describe the circular rotation in our
Galaxy. For this model we have to determine the Galactocentric radius 
of the Sun $R_\odot$ and its circular velocity around the Galactic
centre $\Theta_\odot$. Instead of assuming 
the standard values, $R_\odot = 8.5$~kpc and $\Theta_\odot = 220$~km\,s$^{-1}$,
which are still used in most kinematic distance estimates and Galactic
models \citep[e.g. ][]{cord02}, we carefully investigated the literature
to find the most recent estimates. $R_\odot$ was determined to
be $7.6\pm 0.3$~kpc by \citet{eise05} from a high precision
measurement of the three dimensional orbit of the star S2, which
is orbiting the black hole at the centre of the Galaxy at a distance
of less than $0.2\arcsec$.
This is significantly smaller ($\sim 11$~\%) than the 
previously assumed distance. \citet{feas97} determined the angular velocity 
of circular rotation at the position of the Sun $\Omega_\odot$,
from the Oort constants using data from the Hipparcos satellite. 
$\Omega_\odot = \frac{\Theta_\odot}{R_\odot} = 27.2\pm
0.8$~km~s$^{-1}$~kpc$^{-1}$ gives $\Theta_\odot = 207\pm
10$~km~s$^{-1}$. Another technique to determine $\Theta_\odot$ is the
proper motion of the Galactic Centre. \citet{reid04} found a proper
motion of 6.4\,mas\,yr$^{-1}$ for Sagittarius~A, which translates to a
value of $229\pm 10$~km~s$^{-1}$ for a distance of 7.6~kpc to the
Galactic Centre. Correcting this for the peculiar motion of the Sun,
which \citet{reid04} suggest is between 5 and 12\,km\,s$^{-1}$, we
derive $\Theta_\odot = 220\pm 11$~km~s$^{-1}$.  Both values agree
within their uncertainties, and averaging the two values we obtain
$\Theta_\odot = 214\pm 7$~km~s$^{-1}$. With these values for R$_\odot$
and $\Theta_\odot$ we can derive the systemic velocity
$v_{sys}$ (local standard of
rest velocity) measured from
the Sun in the direction of Wd~1 as a function of distance
(Fig.~\ref{rot}). 

\begin{figure}
   \centerline{\includegraphics[bb = 0 0 520 480,width=9cm,clip]{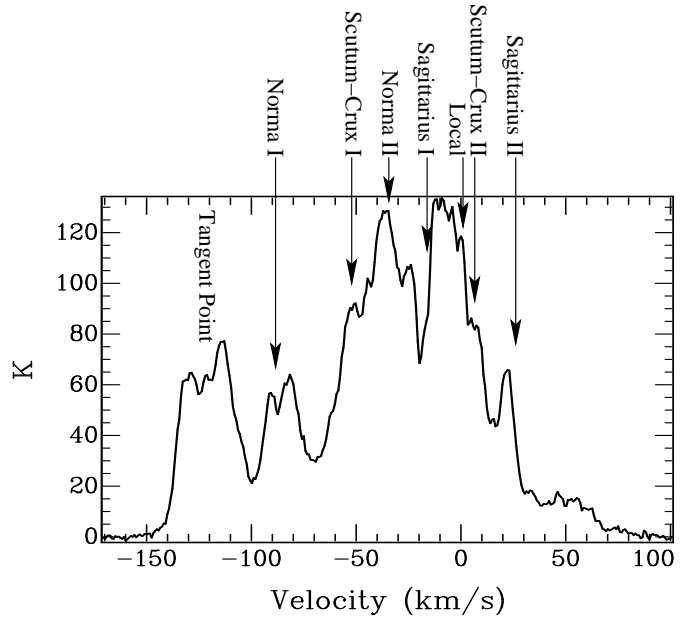}}
   \caption{\ion{H}{i} emission profile averaged over an area of a few
   arcminutes towards Wd~1. The locations of the spiral arms as listed in
   Table~\ref{sparm} are indicated. The Local arm is indicated at 
   0~km\,s$^{-1}$}
   \label{hiemis}
\end{figure}

We tested our assumption of a flat rotation curve by comparing the
measured Tangent point velocity from the SGPS data with the velocity
extrapolated from our values for $\Theta_\odot$.
The Tangent point at any line of sight through the inner Galaxy
is the point at which we look tangential to the circular motion and
hence find the largest line of sight velocity.
For flat rotation we
should observe a radial velocity of $-139$~km~s$^{-1}$ at the Tangent
point in the direction of $l = 339.6^\circ$ (Fig.~\ref{rot}). This
agrees reasonably well with the observed value of somewhere between
$-130$~km~s$^{-1}$ and $-135$~km~s$^{-1}$ (Fig.~\ref{hiemis}). 
This is actually a quite remarkable
result, since the Tangent point in the direction of Wd~1 is at a
distance of about 7~kpc from the Sun at a Galacto-centric radius of
about 2.6~kpc. This should be well within the area which is heavily
influenced by the central bar of the Milky Way. \citet{benj05}
determined that the central bar has a maximum distance of 4.4~kpc from the
Galactic centre, using a Galacto-centric distance of 8.5~kpc for the
Sun.  With the value of $R_\odot = 7.6$~kpc this
would translate to a maximum distance of 3.9~kpc which is
significantly higher than the Galactoc-centric distance of the Tangent
point.

To determine the distance to the spiral arms in the direction of Wd~1
we utilise the distribution of free electrons in our Galaxy as
described in the models of \citet{tayl93} and \citet{cord02}. The
peaks in electron density indicate the locations of the spiral arms in
this direction of the Galaxy (Fig.~\ref{mw}). Since \citet{tayl93} and
\citet{cord02} used a Galacto-centric radius of $R_\odot = 8.5$~kpc
for the Sun in their spiral arm model, we re-scaled this
profile for $R_\odot = 7.6$~kpc. The distances to the nearby
\ion{H}{ii} regions used to define the distance of the nearby spiral arms
were derived spectroscopically. Hence, we left the distance to the two
nearby spiral arms alone and only re-scaled the distances to the other
more distant spiral arms.  With the rotation curve in
Figure~\ref{rot}, we derive the systemic velocities at the centres of
the individual spiral arms.  These newly determined distances and
systemic velocities are listed in Table~\ref{sparm}. A comparison of
these values with the \ion{H}{i} emission profile in the direction of
Wd~1 is shown in Fig.~\ref{hiemis}.

\begin{table}
\caption{The systemic velocity and distance to the spiral arms along the line 
of sight toward Wd~1 determined with the Spiral arm model in Fig~\ref{mw} and
the rotation curve in Fig.~\ref{rot}.}
\label{sparm}
\centering
\begin{tabular}{lcc}
\hline\hline
Spiral Arm & $v_{sys}$ [km\,s$^{-1}$] & $d_{lit}$ [kpc] \\
\hline
Sagittarius I & -17 & 1.5 \\
Scutum-Crux I & -52 & 3.5 \\
Norma I & -89 & 4.9 \\
Norma II & -35 & 11.5 \\
Scutum-Crux II & +6 & 15.0 \\
Sagittarius II & +26 & 18.4 \\
\hline
\end{tabular}
\end{table}

The locations of the individual spiral arms in velocity space agree
remarkably well with the brightness temperature peaks in the
\ion{H}{i} emission profile (Fig.~\ref{hiemis}). The emission peak
between $-$140 and $-$100~km\,s$^{-1}$ near the Tangent point is
produced by a pile up in velocity space rather than a real density
enhancement. In this region, the velocity does not change
significantly with distance, resulting in one velocity channel
containing the emission from \ion{H}{i} distributed over a much larger
distance interval than in other locations along the line of
sight. This creates a peak in the \ion{H}{i} emission profile close to
the Tangent point velocity.  There are two spiral arms that can be
seen distinct from the others, Norma I with a centre velocity of
$-$89~km\,s$^{-1}$ and Sagittarius II, for which the model centre
velocity of +26~km\,s$^{-1}$ is very close to the emission peak at
+22~km\,s$^{-1}$.  Since Sagittarius II is supposed to be well outside
the Solar Circle on the other side of the Galaxy in any Galactic
model, this agreement with the \ion{H}{i} emission profile is
incredible and boosts our confidence in our rotation model. The other
spiral arms are clumped into two groups: Scutum-Crux I and Norma II
merge at a radial velocity of about $-$40~km\,s$^{-1}$ and the Local
arm, Sagittarius I, and Scutum-Crux II are at low negative velocities.

\subsection{\ion{H}{i} absorption}

\begin{figure}
   \centerline{\includegraphics[bb = 80 115 825 765,width=8.0cm,clip]{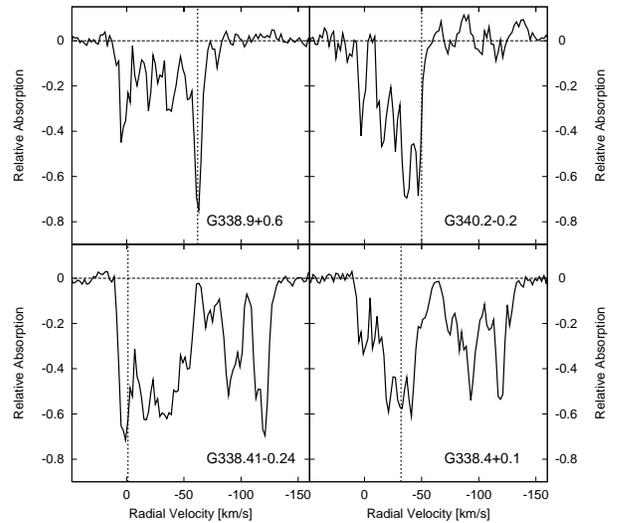}}
   \caption{\ion{H}{i} absorption profiles of four radio bright \ion{H}{ii}
   regions in the vicinity of Wd~1. To make the individual
   components better comparable we plotted the relative absorption --
   the ratio of absorbed emission to total emission -- as a function of radial
   velocity. The dotted vertical lines indicate the systemic velocities of
   the \ion{H}{ii} regions \citep{russ03}, which, within their uncertainties,
   are in excellent agreement with the absorption peaks.}
   \label{abs1}
\end{figure}

\begin{table}
\caption{Characteristics of radio bright \ion{H}{ii} regions and supernova
   remnants in the vicinity of Wd~1. Column 1 contains a name based on the 
   Galactic coordinates of the object. Columns 2 and 3 contain information 
   about the systemic velocity and derived kinematic distance found in the 
   literature \citep{russ03}. Column 4 contains the distance to the source 
   based on the rotation curve and method described in Sections~3.1 and 4.1, 
   respectively. Column 5 indicates the proposed spiral arm the object 
   resides in.}
\label{vsys}
\centering
\begin{tabular}{lcccl}
\hline\hline
& & & & \\
Source & $v_{sys}$ & $d_{lit}$ &
 $d_{new}$ & Spiral Arm \\
  & [km\,s$^{-1}$] & [kpc] & [kpc] & \\
& & & & \\
\hline
& & & & \\
G337.8$-$0.1 & & & $\approx$11 & Norma II \\
G337.95$-$0.48 & $-41$ & 3.1 & $2.9^{+1.2}_{-0.4}$ & Scutum Crux I\\
G338.0$-$0.1 & $-51$ & 12.0 & $10.7^{+1.4}_{-0.4}$ & Norma II \\
G338.41$-$0.24 & $-1$ & 15.7 & $14.0^{+4.5}_{-0.9}$ & Scutum-Crux II \\
G338.4$+$0.1 & $-32$ & 13.1 & $11.7^{+2.0}_{-0.5}$ & Norma II \\
G338.8$+$0.6 & $-62$ & 4.3 & $3.9^{+0.9}_{-0.4}$ & Scutum-Crux I\\
G338.5+0.1 & & & $\approx$11 & Norma II\\
G338.9$-$0.1 & $-38$ & 3.1 & $2.8^{+1.2}_{-0.6}$ & Scutum-Crux I\\
G338.9+0.4 & & & $\approx$3.9 & Scutum-Crux I\\
G339.13$-$0.41 & $-38$ & 3.1 & $2.8^{+1.2}_{-0.6}$ & Scutum-Crux I\\
Wd~1 & $-55$ &  & $3.6^{+1.0}_{-0.4}$ & Scutum-Crux I\\
G339.58$-$0.12 & $-34$ & 2.8 & $2.6^{+1.3}_{-0.7}$ & Scutum-Crux I\\
G339.84$+$0.27 & $-20$ & 14.1 & $12.5^{+2.7}_{-0.6}$ & Norma II \\
G340.2$-$0.2 & $-50$ & 3.7 & $3.5^{+1.0}_{-0.5}$ & Scutum-Crux I\\
G340.24$-$0.48 & $-61$ & 4.4 & $3.9^{+0.9}_{-0.3}$ & Scutum-Crux I\\
G340.6+0.3 & & & $\approx$15 & Scutum-Crux II\\
& & & & \\
\hline
\end{tabular}
\end{table}

The radio continuum emission of the \ion{H}{ii} region produced by the
Wd~1 cluster is not bright enough at 1420~MHz to produce a prominent
\ion{H}{i} absorption profile. Its peak brightness temperature is only
about 24~K in the SGPS data and the source is barely resolved. Its
diameter is about $3\arcmin$ at a resolution of $2\arcmin$. Therefore,
we have to average over several velocity channels to amplify the weak
absorption signal and increase its signal-to-noise ratio. For each
spiral arm along the line of sight we tried to average those channels
that are likely to produce the deepest absorption. In these absorption 
profiles we can identify those velocity
intervals in the spiral arms where the \ion{H}{i} produces deep and
distinct absorption features and those that do not. As a reference we
used the absorption profiles of the nearby \ion{H}{ii} regions shown
in Fig.~\ref{abs1}.  Two of these, representing the G338.9$+$0.6 and
the G340.2$-$0.4 complexes, are Scutum-Crux I objects according to
their systemic velocity (see Table~\ref{vsys}). The other two
\ion{H}{ii} regions are located beyond the Tangent point in Norma II
(G338.4+0.1 complex) and G338.41-0.24 might even reside in Scutum-Crux
II.

\begin{figure}
   \centerline{\includegraphics[bb = 40 90 415 605,width=8.5cm,clip]{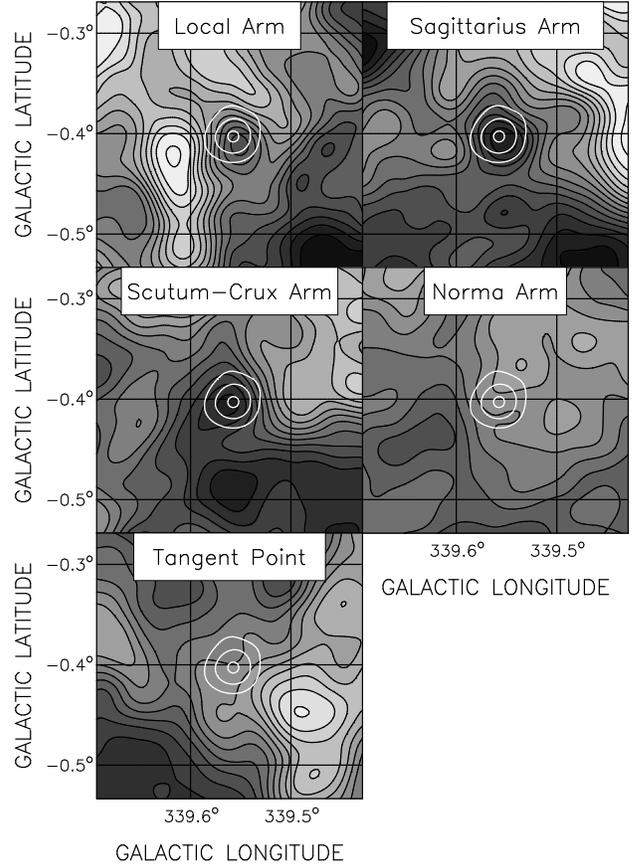}}
   \caption{\ion{H}{i} channels images averaged together to
   demonstrate in which spiral arms Wd~1 is absorbed. For the
   individual arms we averaged over: +5 to $-$1~km\,s$^{-1}$ (Local
   arm), $-$10 to $-$20~km\,s$^{-1}$ (Sagittarius arm), $-$32 to
   $-$45~km\,s$^{-1}$ (Scutum-Crux arm), $-$85 to $-$102~km\,s$^{-1}$
   (Norma arm), and $-$110 to $-$125~km\,s$^{-1}$ (Tangent Point). In 
   these images white denotes high and black weak emission. 
   The radio
   continuum emission of Wd~1 is indicated by the white contours.}
   \label{wd1abs}
\end{figure}

The absorption of the three nearby arms, the Local arm, Sagittarius I,
and Scutum-Crux I, was identified in the absorption profiles of the
two \ion{H}{ii} regions in Scutum-Crux I since their profiles are not
contaminated by absorption of \ion{H}{i} located beyond the Tangent 
point. 
For Norma I and the Tangent point we used the two \ion{H}{ii}
regions that are located beyond the Tangent point
on the other side of the Galaxy.  We found
distinct absorption in the Local arm between +5 and
$-$1\,km\,s$^{-1}$, for Sagittarius I between $-$10 and
$-$20\,km\,s$^{-1}$, for Scutum-Crux I between $-$32 and
$-$45\,km\,s$^{-1}$, for Norma I between $-$85 and
$-$102\,km\,s$^{-1}$, and for the Tangent point between $-$110 and
$-$125\,km\,s$^{-1}$. For Scutum-Crux I we did not use negative
velocities higher than $-$45\,km\,s$^{-1}$ to avoid confusion with a possible
bubble centered at Wd~1 (see section 3.3).

The results of our averaging procedure are shown in
Fig~\ref{wd1abs}. It is apparent that the radio continuum emission
from Wd~1 is absorbed by the Local arm, Sagittarius I, and Scutum-Crux
I. This is indicated by a hole in the distribution of
\ion{H}{i} emission at the position
of Wd~1. There is no evidence for absorption in Norma I or the Tangent
point gas. This indicates a location in Scutum-Crux I or at the near
edge of Norma I. The latter, however, is rather unlikely, because
there must not be any absorbing Norma I material between us and
Wd~1. All \ion{H}{ii} regions shown in Fig.~\ref{abs1} have a very
deep distinct absorption feature at their systemic velocity, which is
likely created by material in their vicinity in the cloud complex from
which the stars in those \ion{H}{ii} regions were formed. Since the
remains of these clouds are expected to be very dense and cold the
optical depth should be rather high.  This should produce deep
absorption lines. If Wd~1 is located in the Norma arm it cannot
have such a component. This makes a Scutum-Crux arm location more 
likely, however, it does not entirely exclude the possibility of a location
at the near edge of the Norma arm.

\subsection{The neutral environment of Wd~1}

\begin{figure*}
   \centerline{\includegraphics[bb = 50 110 580 630,width=11cm,clip]{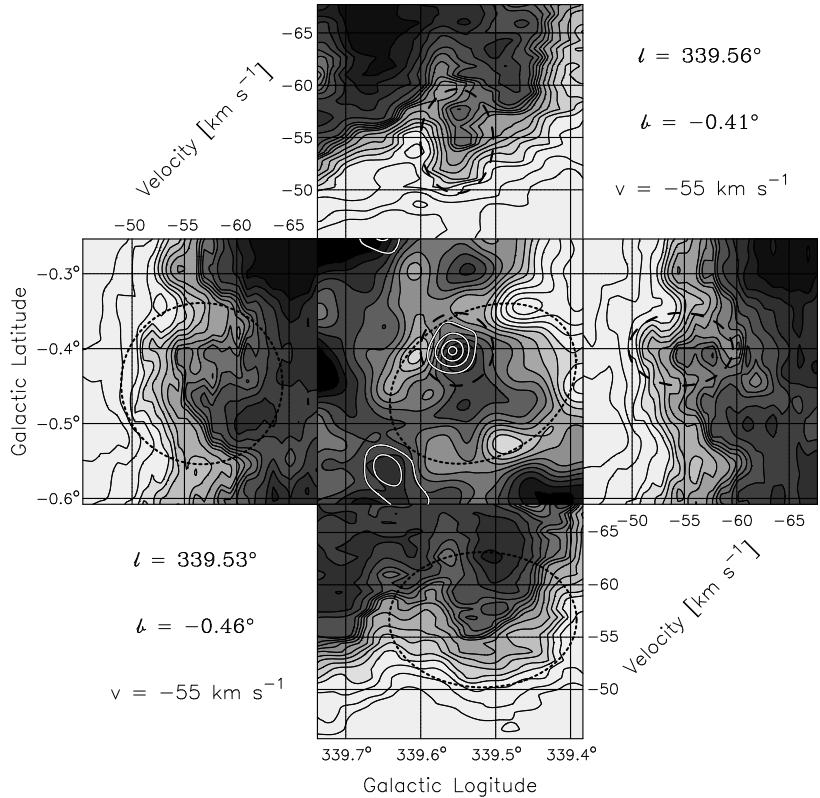}}
   \caption{Three views of the two small expanding bubbles, B1 and B2,
   around Wd~1, indicated by dashed and dotted ellipses
   respectively. Longitude-velocity slices are displayed at the top
   and bottom and the velocity-latitude slices in the left and right
   panels. The HI channel map at about -55 km/s in longitude-latitude
   is shown in the centre, with Wd~1 indicated by the white
   contours. The bright emission in the velocity slices mark the gas
   at the far edge of Scutum-Crux I. In these images black represents
   low and white high intensity.}
   \label{smbub}
\end{figure*}

We carefully investigated the HI data in the velocity range of the
Scutum-Crux arm and at velocities corresponding to the near edge of
the Norma arm.  We found one prominent feature in velocity space which
is exactly centered at the position of Wd~1 at a velocity of -55
km\,s$^{-1}$. This feature is shown in Fig~\ref{smbub}, indicated by
the dashed ellipses (top and right panels).  The dashed circle in the
centre image marks the location of this expanding bubble projected
back to the map plane. This feature, B1, cannot be an absorption
feature since the depression visible at the location of Wd~1 is deeper
than the peak brightness of Wd~1's radio continuum emission.  In the
absorption map of the Scutum-Crux arm (Fig.~\ref{wd1abs}) the velocity
interval related to this expanding bubble was omitted to avoid
confusion with a hole in the \ion{H}{i} map that is created by a lack
of \ion{H}{i} and not actual \ion{H}{i} absorption.

Feature B1 seems to be located at the edge of a much larger elliptical
bubble (B2), indicated by the dotted ellipses in
Fig.~\ref{smbub}. This feature is very obvious in the map plane but
is not as clear in velocity space (left and bottom panels of
Fig.~\ref{smbub}). There are fingers of emission emerging from the
bright Scutum-Crux I gas towards higher negative velocities,
indicating an expansion for this feature. However, these fingers are
not closed in an end cap, which would mark the part of the bubble that
is moving towards us. Bubble B1 is closed in velocity space, which
makes it very easy to detect.  The expansion velocity of B1 is
5~km\,s$^{-1}$, as readily seen in Fig.~\ref{smbub}.  The expansion
velocity of B2 is not easy to identify since the end cap is missing,
but the velocity fingers seem to indicate a somewhat higher expansion
velocity. The spatial coincidence of the two features suggests both
are associated with Wd~1, supported by their central velocity of
$-$55~km\,s$^{-1}$. This velocity is well within the velocity 
interval predicted for Wd~1 by the \ion{H}{i} absorption measurements. 
Therefore we adopt a radial velocity
of $-55\pm 3$~km\,s$^{-1}$ for Wd~1.  Uncertainties for the radial
velocities along the line of sight are discussed in Sect. 4.1.

\begin{figure}
   \centerline{\includegraphics[bb = 50 60 540 535,width=8cm,clip]{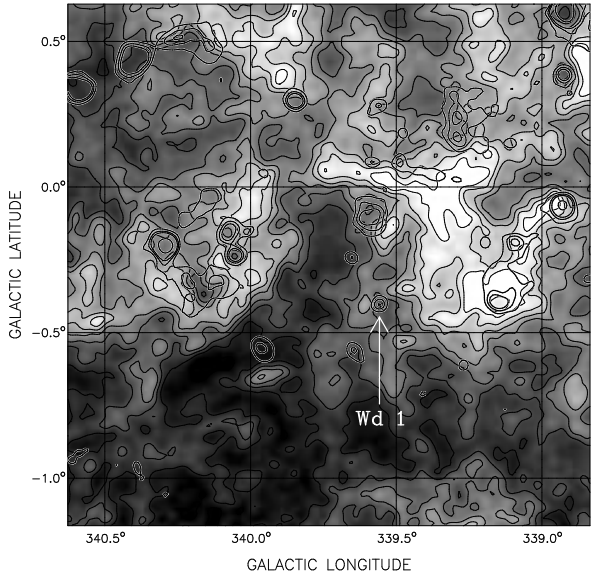}}
   \caption{\ion{H}{i} channelmap at a radial velocity of
   $-$55~km\,s$^{-1}$ from the SGPS. Black contours go from 20~K to
   90~K in steps of 10~K. Grayish contours indicate the 1420~MHz
   continuum emission. The location of Wd~1 is indicated.}  
\label{bb}
\end{figure}

At a radial velocity of $-$55~km\,s$^{-1}$ a larger field of view
reveals a much larger bubble-like feature (B3), which is open to the
south, away from the Galactic plane (Fig.~\ref{bb}). The emission
associated with Bubble B2 can be seen in the lower right corner of
B3. To the north, B3 consists of a shell with two large, bright, and
complex emission regions to the east and west. The eastern region contains the
\ion{H}{ii} complex G340.2$-$0.2, with a radial velocity of
$-$50~km\,s$^{-1}$ (Table~\ref{vsys}). This suggests that Wd~1 and the
\ion{H}{ii} region complex G340.2$-$0.2 are evolving in the same
environment, at the far side of Scutum-Crux I.

\section{Discussion}

\subsection{The distance to Westerlund 1}

We determine a radial velocity of $-55\pm 3$~km~s$^{-1}$ for Wd~1 from
the central velocity of the bubbles in which we believe it is located.
Using the rotation curve determined in Sect. 3.1 (see Fig.~\ref{rot})
this translates to a distance of $3.6\pm 0.2$~kpc or $10.6\pm
0.2$~kpc.  The closer distance is preferable, because the \ion{H}{ii}
region of Wd~1 does not indicate any absorption in the \ion{H}{i} gas
of Norma I or near the Tangent point. Additionally, the previous
distance estimates are $\sim4-5$~kpc. A comparison with the spiral arm
model in Fig.~\ref{mw} and the calculated distances to the spiral arms
in Table~\ref{sparm} reveals that Wd~1 is located in the Scutum-Crux I
arm.

A possible source of uncertainty for our rotation curve could be the
presence of non-circular motion. Non-circular velocity components due
to streaming motion can reach a maximum of about $\pm 10$~km~s$^{-1}$
\citep{burt72}. This would result in $\pm 8$~km~s$^{-1}$ projected to
the line of sight towards Wd~1 assuming a distance of 3.6~kpc. Another
source of non-circular motion could be a spiral shock, which is always
directed towards the Galactic centre. At the location of Wd~1 the
spiral shock would be directed away from us, leading to an
underestimate of the actual distance. Typical values for the velocity
shift due to the spiral shock could reach a maximum of 30~km~s$^{-1}$
\citep{fost06}, which would translate to about $24$~km~s$^{-1}$
projected to the line of sight. Random motion add another $\pm
5$~km~s$^{-1}$ to our uncertainty.  Taking all these uncertainties
into account, we derive a systemic velocity of
$-55^{+9}_{-26}$~km~s$^{-1}$ for Wd~1. In the direction of Wd~1,the
velocity versus distance gradient is rather steep (see Fig~\ref{rot}),
so the large velocity uncertainty does not translate to a large
distance uncertainty. This radial velocity corresponds to a distance of
$3.6^{ +1.0}_{-0.4}$~kpc, or by choosing the centre of the uncertainty
interval as the most probable distance, we get $3.9\pm0.7$~kpc. This
result agrees very well with the latest independent distance estimates
of $4.0\pm 0.3$~kpc \citep{bran05} and $4.7\pm 1.1$~kpc
\citep{crow06}. This distance is consistent with a location at the far
side of the Scutum-Crux I arm. Furthermore, the bubbles B1 and
B2 in Fig.~\ref{smbub} seem to be emerging from the bright gas at the
high velocity edge of the Scutum-Crux I arm towards higher negative
velocities, supporting a location on the far side of this arm.

\subsection{Distance estimates for \ion{H}{ii} regions and SNRs in the 
vicinity of Wd~1}

\begin{figure}
   \centerline{\includegraphics[bb = 60 60 550 310,width=8.5cm,clip]{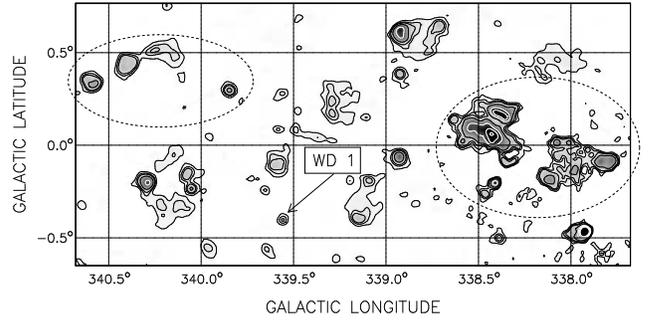}}
   \caption{Radio continuum emission at 1420~MHz around Wd~1 taken from the
   end channels of the SGPS \ion{H}{i} observations. Contours are at 11, 17, 23, and
   40\,K (black) and 80, 120, and 160\,K (white). The dotted ellipses enclose
   SNRs and \ion{H}{ii} regions that are believed to be beyond the Tangent 
   point. The location of Wd~1 is indicated.}
   \label{tp21}
\end{figure}

\begin{figure}
   \centerline{\includegraphics[bb = 80 115 825 765,width=8.0cm,clip]{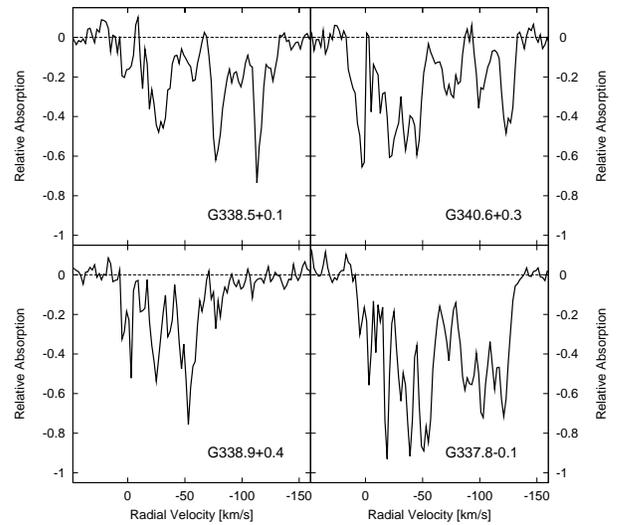}}
   \caption{\ion{H}{i} absorption profiles of one radio bright
   \ion{H}{ii} region (G338.9+0.4) and three bright SNRs in the
   vicinity of Wd~1.}
   \label{abs2}
\end{figure}

In Figure~\ref{tp21} we show a 1420-MHz radio continuum image of the
area around Wd~1, created from the end-channels of the \ion{H}{i} data
that do not contain any neutral hydrogen emission. Wd~1 is located in
a rich area of radio bright \ion{H}{ii} regions and supernova
remnants.  In Table~\ref{vsys} we list \ion{H}{ii} regions and
complexes for which the systemic velocity is known and for which the
distance ambiguity in the inner Galaxy, where each radial velocity
corresponds to two possible distances, was solved \citep{russ03}. With
the rotation curve shown in Fig.~\ref{rot}, new distances to these
objects can be deduced. A comparison of the systemic velocities of
these \ion{H}{ii} regions with the centre velocity of each spiral arm,
determined in Sect. 3.1 (see Table~\ref{sparm}), gives the most likely
spiral arm where these objects reside (Table~\ref{vsys}).  Almost all of
these sources are apparently concentrated in the Scutum-Crux I arm and
the Norma II arm. Only G$338.41-0.24$ seems to be located even further
away in the Scutum-Crux II arm. Objects believed to be
beyond the Tangent point are encircled by dashed ellipses in
Fig.~\ref{tp21}.  It is quite curious that no objects were found in
either Sagittarius I or Norma I arms

\ion{H}{i} absorption profiles for four additional radio bright
sources can also be determined (Fig.~\ref{abs2}). The distances and
systemic velocities of these objects were previously unknown.  These
absorption profiles were compared with those in Fig~\ref{abs1} to
determine the most likely spiral arm location for these objects. We
then assume the distance to the centre of that particular spiral arm
to be the most likely value for these objects, with uncertainties
$\sim\pm1$~kpc (Table~\ref{vsys}).

The absorption profile of the HII region G338.9+0.4 in Fig.~\ref{abs2}
is very similar to G$338.9+0.6$ and G$340.2-0.2$ (see Fig.~\ref{abs1})
with no evidence of absorption in Norma I or at the Tangent point.
This suggests G$338.9+0.4$ could be a part of the G$338.9+0.6$
\ion{H}{ii} region complex, just to the north (see
Fig~\ref{tp21}). Therefore, we propose a Scutum-Crux I location at a
distance of about 3.9~kpc.

The SNR G340.6+0.3 is absorbed by Norma I and the Tangent point gas
and also shows a deep absorption feature with a relative absorption of
almost 0.7 at about 0~km\,s$^{-1}$ (Fig~\ref{abs2}). This makes its
absorption profile very similar to that of G$338.41-0.24$
(Fig~\ref{abs1}).  No other source shows such a deep and wide
absorption feature around a velocity of 0~km\,s$^{-1}$. This indicates
a location on the other side of the Galaxy in Scutum-Crux II at about
15~kpc. This gives the SNR a diameter of about 26~pc. It is not
unusual that absorption features that are far away seem to be deeper
than those that are nearby. The reason for this is simply the
area probed by the observing beam becomes bigger with distance so that
more \ion{H}{i} is detected within one beam.

The SNRs G337.8$-$0.1 and G338.5+0.1 show absorption by Norma I and at
the Tangent point but lack the deep and wide absorption feature at
0~km~s$^{-1}$.  Their absorption profiles look very similar to that of
G338.4+0.1 in Fig.~\ref{abs1}. Therefore these two SNRs are most
likely Norma II objects at a distance of about 11~kpc. This gives
G338.5+0.1 a diameter of about 30~pc and G337.8$-$0.1 an extent of
about 30$\times$20~pc.

\subsection{The Wd~1 \ion{H}{ii} region}

\begin{figure}
   \centerline{\includegraphics[bb = 55 45 525 495,width=7cm,clip]{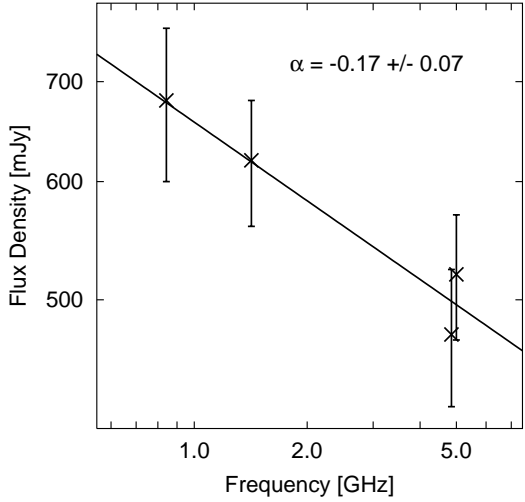}}
   \caption{Radio continuum spectrum of the \ion{H}{ii} region,
   produced by the members of Wd~1.}
   \label{spec}
\end{figure}

We determined radio continuum flux densities for the Wd~1 \ion{H}{ii}
region at 1.4~GHz ($620\pm 60$~mJy) from the continuum end channels of
the SGPS \ion{H}{i} data and at 843~MHz ($680\pm 80$~mJy) from data of
the Molonglo Survey \citep{gree99}. We added two flux density
measurements from the literature at 5~GHz \citep[520~mJy, ][]{hayn79} and at
4.85~GHz \citep[474~mJy, ][]{wrig94} (Fig.~\ref{spec}). Since no
error was given for the \citet{hayn79} value we weighted both fluxes
equally. The error bars plotted in Fig.~\ref{spec} represent 10\% 
of the flux density.
A weighted least-squares fit gives a
spectral index of $\alpha = -0.17\pm 0.07$, that we interpret as
arising from optically thin thermal plasma.

The observed diameter of Wd~1 in our 1.4~GHz data is $3\farcm5$ at a
resolution of $2\arcmin$. This results in an actual diameter of
$3\arcmin$.  This agrees with its appearance in the 843~MHz Molonglo
Survey \citep{gree99}. The observed peak of the radio continuum
emission is about 24~K. Correcting for the synthesised beam, the
\ion{H}{ii} region has a peak brightness temperature of 33~K at
1.4~GHz. The observed brightness temperature $T_B$ at frequency $\nu$
is related to the electron temperature $T_e$ inside the \ion{H}{ii}
region and the optical depth $\tau$ by:
\begin{equation}
T_b(\nu) = T_e\,(1 - e^{-\tau(\nu)}),
\end{equation}
where the optical depth at frequency $\nu$ is given by:
\begin{equation}
\tau(\nu) = 8.235\,10^{-2}\,\left(\frac{T_e}{\rm K}\right)^{-1.35}\,
\left(\frac{\nu}{\rm GHz}\right)^{-2.1}\,\left(\frac{EM}{\rm pc\, cm^{-6}}
\right).
\end{equation}
\citep[e.g.][]{rohl04}. Here $EM$ is the emission measure related to
the path length $l$ through the \ion{H}{ii} region and the electron density 
$n_e$ inside it by:
\begin{equation}
EM = \int n_e^2\,dl
\end{equation}
With an electron temperature of $T_e = 8000\pm 2000$~K and a
path length of 3.5~pc ($3\arcmin$ at 3.9~kpc) through the \ion{H}{ii}
region, we estimate an optical depth of $0.0041^{+0.0014}_{-0.0008}$
and an emission measure of $19000^{+8000}_{-6000}$\,pc\,cm$^{-6}$.
Hence, $n_e = 74\pm 13$\,cm$^{-3}$ and the total mass of the ionized
gas inside the \ion{H}{ii} region is $53\pm 9$\,M$_\odot$.  The
frequency at which this emission becomes optically thick (i.e.
$\tau_\nu=1$) is $\approx 100$~MHz.

Alternatively, the electron density of the \ion{H}{ii} region can be 
calculated from the observed flux density. For 
an optically thin Maxwellian plasma, the flux $S_\nu$ is given by:
\begin{equation}
S_\nu = 5.7\,10^{-56}\,T_e^{-0.5}\,g_{ff}\,E_V\,d^{-2}~{\rm mJy}, 
\end{equation}
where $g_{ff}$ is the Gaunt-factor defined by:
\begin{equation}
g_{ff} = \frac{\sqrt{3}}{\pi}\,\left(17.7 + ln\left(\frac{T_e^{1.5}}{\nu}\right)\right)
\end{equation}
and $E_V$ is the volume emissivity:
\begin{equation}
E_V = \int n_e^2 dV.
\end{equation}
The radio spectrum in Fig.~\ref{spec} indicates that the emission is
optically thin over the entire observed frequency range.  Using the
flux density at 843~MHz, we derive an electron density of $n_e = 65\pm
7$\,cm$^{-3}$ and a total ionized mass of $M = 47\pm 5$\,M$_\odot$.
This is in excellent agreement with the value derived from the
observed brightness temperature.

\subsection{Wd~1 and its neutral environment}

In Section 3, two small (B1 and B2) and one large (B3) bubble in the
\ion{H}{i} data were described. We believe these are related to Wd~1.
The determination of the mass of atomic material in these bubbles is
difficult since these objects, in particular bubble B3, seems to
merge with the surrounding ISM in the \ion{H}{i} images (
Figs.~\ref{smbub} and \ref{bb}). Bubble B1 and B2 together contain a 
mass of about
300~M$_\odot$ with an error of at least 50~\%, assuming the \ion{H}{i}
is optically thin and the mass ratio between hydrogen and helium is
10:3. The inner edge of bubble B1 seems to be just outside the Wd~1
\ion{H}{ii} region (see Fig.~\ref{smbub}) giving a diameter of about
5~pc at a distance of 3.9~kpc. At the same distance, bubble B2 has an
extent of about 18$\times$10~pc. \citet{mccl02} defined the dynamic
age of a stellar wind bubble that expands as $\propto t^{0.3}$ by:
\begin{equation}
t_6 = 0.29\,\frac{R}{v_{exp}},
\end{equation}
where $t_6$ is the dynamic age of the bubble in Myr, $R$ the radius
in pc, and $v_{exp}$ the expansion velocity in km\,s$^{-1}$.  For
bubble B2 the maximum expansion radius calculated from the position of
Wd~1 is about 10~pc. If we use an expansion velocity of 5~km\,s$^{-1}$
for B2, determined for bubble B1, we derive a dynamic age of about
600,000~yr for B2. Since its structure in velocity space (see
Fig.~\ref{smbub}) implies a somewhat higher expansion velocity than
that for bubble B1 and the radius is a maximum expansion radius, this
age represents an upper limit.

\citet{crow06} determined an age of $4.5-5.0$~Myrs for Wd~1, based on
the ratio of WR stars to red and yellow hypergiants.  Clearly,
bubbles B1 and B2 cannot be the stellar wind bubble produced in the
early evolution of Wd~1; they are simply too young. Both B1 and B2 could
have been created after the last supernova explosion pushed away all
material inside the stellar wind bubble. The recent discovery of the
X-ray pulsar CXO J164710.2-455216 inside Wd~1 implies that at least
one supernova explosion occured inside Wd~1 \citep{muno06}. In this
case we speculate that B1 and B2 originate from the only source of
matter and energy left after the last supernova occurred, the combined 
winds and mass loss of the cluster stars. In that
scenario, the gas in the shells of those bubbles must consist mostly
of recombined material previously ejected by the stars in their winds,
since the supernova shock wave should have removed any remaining
material in the ambient medium. This would imply that the last
supernova explosion happened less than 600,000~yr ago. Additionally, 
the ionized material inside
the \ion{H}{ii} region would also be material ejected by the stars
in their winds. The total mass of the ionized gas
plus the B2 \ion{H}{i} shell is $\sim350$~M$_\odot$. If we assume
these structures are 500,000~yr old we require 70 stars with an
average mass-loss rate of $10^{-5}$\,M$_\odot$\,yr$^{-1}$ for each star, 
which is highly feasible 
for the stellar population in Wd~1.

We suspect that the impact of the original stellar wind bubble, created
by the members of Wd~1 in their early stage of evolution, is
represented by bubble B3 (Fig.~\ref{bb}). 
Its diameter is about 50~pc at a distance of
3.9~kpc. The location of Wd~1 in the lower right corner of this
feature is likely the result of a highly structured environment. To
the South, away from the Galactic plane, the density is expected to
decrease in which case the winds of the members of Wd~1 pushed the
material away from the Galactic plane into the Galactic halo.
To the West we find a dense cloud complex that
impeded free expansion, whereas to the East the winds
of the stars pushed material into the cloud centered at the
G$340.2-0.2$ \ion{H}{ii} region complex. In Fig.~\ref{tp21}
we can identify three bright compact radio sources that belong
to this complex at positions (l,b): (340.3,-0.2), (340.1,-0.15), and
(340.1,-0.25). These are all sources with far infrared colours
characteristic of ultra-compact \ion{H}{ii} regions, which are
known to be regions of OB star formation \citep{bron96}.
This indicates that the G340.2-0.2 \ion{H}{ii} region complex is 
younger than Wd~1 and its formation may well have been triggered
by the interaction of the early Wd~1 stellar wind or a
supernova explosion with the clouds in which G340.2-0.2 is embedded.

To the North of bubble B3 we find a shell
that might still be expanding. Since this shell is the only dynamic
feature we observe, we use its distance of 45~pc from Wd~1 as the
expansion radius.  
We do not see any expansion of B3 in the velocity slices of our
\ion{H}{i} data. If we assume that the dynamic age of bubble B3 equals
the assumed age of Wd~1, an expansion velocity of about 3~km\,s$^{-1}$
is implied, which is a reasonable value. Bubble B3 is certainly 
much older than B2 and B1, because it is many times as large and
a low velocity is indicated by the absence of any visible feature
in the velocity slices.
Unfortunately we cannot estimate the mass of the bubble because of the
confusion with the emission regions to the east and the west.

\section{Summary}
Based on \ion{H}{i} observations from the SGPS, we have established
that Wd~1 is in the Scutum-Crux arm of the Galaxy. We find three large
bubbles in which Wd~1 is located at a radial velocity of
$-55^{+9}_{-26}$~km\,s$^{-1}$.  Using a flat rotation model of the Galaxy and
adopting a galacto-centric distance of 7.6~kpc \citep{eise05} and a
velocity of $214\pm7$~km\,s$^{-1}$ for the Sun around the Galactic
Centre, we transpose the radial velocity of the bubble features to a
distance of $3.9\pm0.7$~kpc at a Galacto-centric radius of about
4.2~kpc. We are confident in our derived distance-velocity calibration 
in this direction, since it predicts to within
$\pm4$~km\,s$^{-1}$ the velocity of the Tangent point and the velocity
of the \ion{H}{i} gas in the Sagittarius Arm outside the Solar Circle
on the far side of the Galactic Centre. Moreover, this is very
encouraging since it is believed that the position of Wd~1 is in the
region of the Galactic bar. 

A distance of $3.9\pm0.7$~kpc is somewhat less than the earlier
estimate of $4.7\pm1.1$~kpc based on the absolute magnitude
calibration of WR stars \citep{crow06}, but very similar to the
$4.0\pm0.3$~kpc determined by \cite{bran05} from initial analysis of
the photometric detection of the main and pre-main sequence
populations in Wd~1.  Though the uncertainty on these estimates is
relatively large ($\sim10-25\%$), a weighted average of these values
gives us confidence that the distance to Wd~1 is $\sim4$~kpc. Until
the final analysis of the photometric observations described in
\cite{bran05} is available, we argue that this represents the best
distance estimate to Wd~1.

A study of the \ion{H}{i} images from the vicinity of Wd~1 revealed
three bubble-like features B1, B2, and B3. We believe that bubbles
B1 and B2 are the result of the stellar winds of the
members of Wd~1 and contain recombined wind material. We argue bubble
B3 represents the stellar wind bubble created in the early
history of Wd~1. The formation of the \ion{H}{ii} region complex 
G340.2$-$0.2 which is embedded in the eastern edge of B3 may have
been triggered by the interaction of the Wd~1 wind with the dense
clouds in this area. If this is true, the Wd~1 G340.2$-$0.2 pair would
be a nice example of sequential star formation.

\begin{acknowledgements}
We wish to thank Naomi McClure-Griffiths for her help with
the SGPS HI and continuum data. We also wish to thank Tom
Landecker for careful reading of the manuscript.
The Dominion Radio Astrophysical Observatory is a National
Facility operated by the National Research Council Canada.
\end{acknowledgements}

\bibliographystyle{aa}
\bibliography{wd1}

\end{document}